\documentclass{article}
\usepackage{spconf}
\usepackage{amsmath}
\usepackage{amssymb}
\usepackage{graphicx}
\usepackage{gensymb}
\usepackage[colorlinks=true, allcolors=blue]{hyperref}
\usepackage{bbold}
\usepackage{tikz}
\usepackage{array}
\usepackage{doi}
\usepackage{booktabs}
\usepackage{soul}

\newcommand{\x}{{\bf x}}
\newcommand{\y}{{\bf y}}
\renewcommand{\u}{{\bf u}}
\renewcommand{\v}{{\bf v}}

\title{Earthmover-based manifold learning for analyzing molecular conformation spaces}

\name{
    Nathan Zelesko\(^{\spadesuit}\) \qquad
    Amit Moscovich\(^{\diamondsuit}\) \qquad\quad
    Joe Kileel\(^{\diamondsuit}\) \qquad\qquad
    Amit Singer\(^{\diamondsuit, \, \clubsuit}\)
}
\address{
    $^{\spadesuit}$ Department of Mathematics, Brown University\\
    $^{\diamondsuit}$ Program in Applied and Computational Mathematics, Princeton University\\
    $^{\clubsuit}$ Department of Mathematics, Princeton University\\\\
}

\begin{document}

\maketitle
\begin{abstract}
    %100-150 words
In this paper, we propose a novel approach for manifold learning that combines the Earthmover's distance (EMD) with the diffusion maps method for dimensionality reduction.  We demonstrate the potential benefits of this approach for learning shape spaces of proteins and other flexible macromolecules using a simulated dataset of 3-D density maps that mimic the non-uniform rotary motion of ATP synthase.  Our results show that EMD-based diffusion maps require far fewer samples to recover the intrinsic geometry than the standard diffusion maps algorithm that is based on the Euclidean distance.  To reduce the computational burden of calculating the EMD for all volume pairs, we employ a wavelet-based approximation to the EMD which reduces the computation of the pairwise EMDs to a computation of pairwise weighted-$\ell_1$ distances between wavelet coefficient vectors.
\end{abstract}

\section{Introduction}

Proteins and other macromolecules are elastic structures that may deform in various ways.
Since the  spatial conformation of an organic molecule is known to play a key role in its biological function, the complete
description of a molecule must include more than just a single static structure (as is traditionally produced by X-ray crystallography).
Ideally, we would like to map the entire space of molecular conformations.
However, understanding the topology and geometry of these conformation spaces remains one of the grand challenges in the field of structural biology \cite{Frank2018}.

One promising approach is to employ cryo-electron microscopy (cryo-EM) as a tool for structure determination in the presence of conformational heterogeneity \cite{SorzanoEtal2019}.
In cryo-EM, multiple images of a particular macromolecule are taken by a transmission electron microscope and then processed using specialized algorithms.
Traditionally, these algorithms construct an estimate of the mean molecular volume, in the form of a 3-D electrostatic density map.
In particular, this process averages out any  variability in the spatial conformations of the molecules in the sample.
Recent works have applied techniques from the field of manifold learning to cryo-EM data sets, obtaining a low-dimensional representation of the molecular conformation space \cite{SchwanderFungOurmazd2014, MoscovichHaleviAndenSinger2019}.
Specifically, these works build affinity graphs based on the Euclidean distances between molecular volumes (or projection images) and then compute diffusion map embeddings \cite{BelkinNiyogi2003,CoifmanEtal2005}.

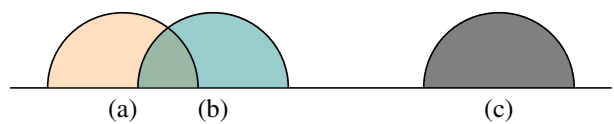
\begin{figure}
    \centering
    \begin{tikzpicture}[semithick]
        \draw (0,0)--(8,0);
        \begin{scope}
            \clip (0,0) rectangle (8,1.02);
            \draw[fill=orange, opacity=0.25] (1.5,0) circle(1);
            \draw[fill=teal, opacity=0.4] (2.7,0) circle(1);
            \draw[fill=black, opacity=0.5] (6.5, 0) circle(1);
            \draw (1.5,0) circle(1);
            \draw (2.7,0) circle(1);
            \draw (6.5, 0) circle(1);
        \end{scope}
        \node at (1.5, -0.3) {(a)};
        \node at (2.7, -0.3) {(b)};
        \node at (6.5, -0.3) {(c)};
    \end{tikzpicture} 
    \caption{\textit{EMD vs. Euclidean distance for translational motion.}  
    Euclidean (or $\ell_2$) distance is only meaningful for measuring small displacements. e.g., the $\ell_2$ distance between half-disks (c) and (a) is the same as between (c) and (b).  By contrast, for any translational motion, the EMD is its magnitude.}
    \label{fig:disks}
\end{figure}

However, the Euclidean distance is suboptimal for capturing the distance between geometric conformations.
Consider, for example, two conformations of a molecule that has only a single moving part.
If the two conformations are distant, the support of the moving part in the two volumes may not intersect, rendering the Euclidean distance independent of the conformational distance. See Fig.~\ref{fig:disks}.
In such cases, in order to apply manifold learning based on a Euclidean metric, one need a dense cover of the conformation space by the molecules in the sample.  Since the number of points in such a cover scales exponentially in the dimension, it may be infeasible to apply these methods, even using the largest existing experimental datasets, which consist of about $n \approx 10^6$ samples.

In this paper, we propose to use the \emph{Earthmover's distance} (EMD), also known as the \emph{Wasserstein metric}, instead of the commonly used Euclidean distance as input to manifold embedding algorithms.
EMD has an intuitive geometric meaning:
it measures the minimal amount of ``work'' needed to transform one pile of mass into another pile of equal mass, where ``work'' is defined as the amount of mass moved times the distance by which it is moved.
In particular, EMD provides a distance metric that is meaningful even between spatial conformations that are far from each other.
Following the discussion above, this property should reduce the number of samples needed to learn the intrinsic manifold.

Methods for computing the EMD, based on off-the-shelf linear programming solvers, are expensive when the number of voxels is large.  Therefore we used a fast approximation to the EMD, based on a wavelet representation \cite{ShirdhonkarJacobs2008}.

\begin{figure}
    \centering
    \begin{tikzpicture}
        \node (ATPpartial) at (0,0) {\includegraphics[width=0.198\linewidth]{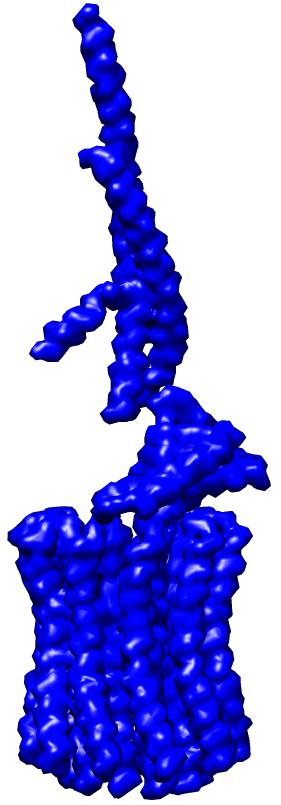}};  
        \node (NoisySlice) at (5.6,0.0) {\includegraphics[height=2.4in]{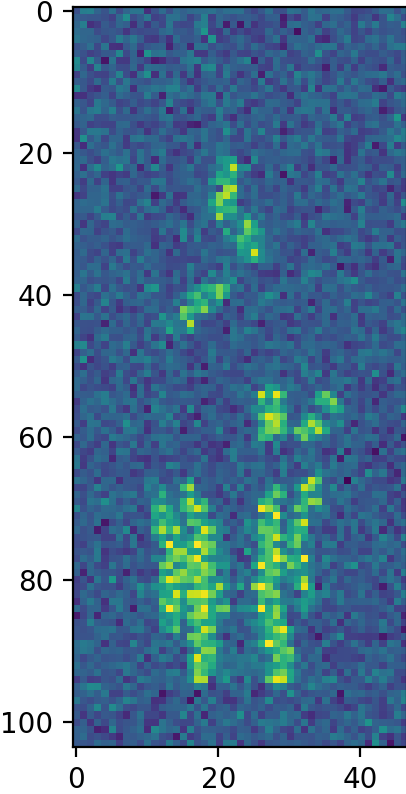}};
        \node (ATPall) at (2.5,0.45) {\includegraphics[width=0.36\linewidth]{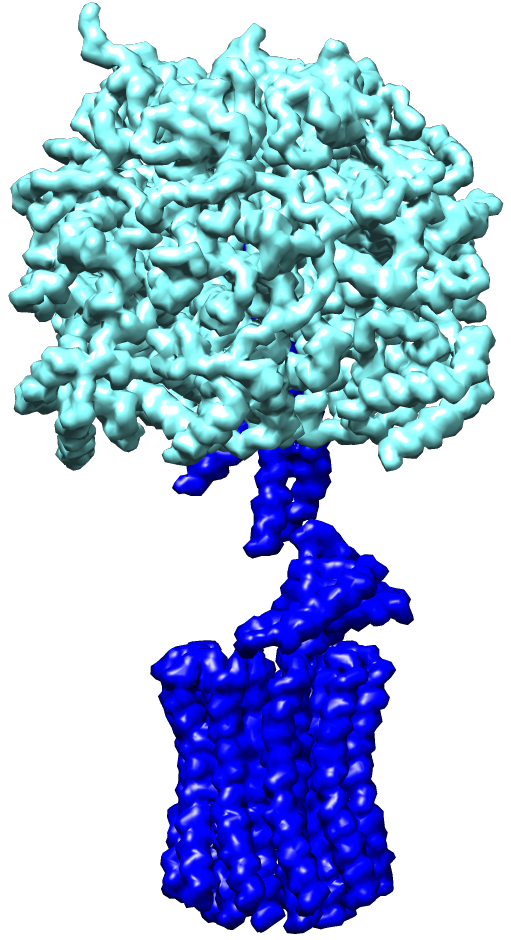}};
        \draw[x=.25cm,y=0.60cm,line width=.2ex,stealth-,rotate=+90] (11,0.1)  arc (-150:150:1 and 1);
    \end{tikzpicture}
    \caption{\textit{ATP synthase.} (left) F$_0$ and axle subunits. They rotate together in the presence of hydrogen ions, forming a tiny electric motor; (middle) the  F$_1$ subunit (in cyan) envelops the axle. As the axle rotates, this subunit assembles ATP; (right) sample slice of the rotated F$_0$ and axle subunits with the additive Gaussian noise. }
    \label{fig:atp_synthase}
\end{figure}

To test our proposal, we compared the standard $\ell_2$-based diffusion maps to EMD-based diffusion maps on a synthetic dataset mimicking the motion of ATP synthase (Fig.~\ref{fig:atp_synthase}).  This dataset samples the underlying manifold in a non-uniform manner since ATP synthase has three dominant conformations that are 120$\degree$ apart \cite{YoshidaMuneyukiHisabori2001}.
The approximate EMD-based approach yields a marked improvement in the number of samples required for learning the conformational manifold, while still offering a computationally feasible algorithm.

\section{Methods}

In this section, we review the basic techniques that underlie Earthmover-based manifold learning.
Our current focus is on learning shape spaces of 3-D volumes, but the same techniques may also be applied to analyze other types of datasets, such as 2-D image sets, 1-D histograms, etc.
To start, let $X = \{\x_1, \ldots, \x_n\}$ be a set of 3-D voxel arrays in $\mathbb{R}^{L^3}$.
We assume that $X$ obeys the \textit{manifold hypothesis} \cite{SorzanoEtal2019,MoscovichJaffeNadler2017,Lee2003}, i.e., $\x_1, \ldots, \x_n$ form a (noisy) sample of a low-dimensional manifold $\mathcal{M} \subset \mathbb{R}^{L^3}$. Our task is to reorganize the data to better reflect the intrinsic geometry of $\mathcal{M}$.

For Riemannian manifolds, eigenfunctions of the \textit{Laplace-Beltrami operator} provide an intrinsic coordinate system \cite{BerardBessonGallot1994,JonesMaggioniSchul2008}.
Accordingly, several popular methods for dimensionality reduction and data representation methods are based on mapping input points using empirical estimates of Laplacian eigenfunctions \cite{CoifmanEtal2005,BelkinNiyogi2003}.
Under the manifold hypothesis, these estimates converge to eigenfunctions of the Laplace-Beltrami operator, or more generally to eigenfunctions of a weighted Laplacian, depending on the construction \cite{TingHuangJordanICML2010}.

We now describe the diffusion maps method \cite{CoifmanEtal2005}.
Let $w:\mathbb{R}^{L^3} \times \mathbb{R}^{L^3} \rightarrow \mathbb{R}$ denote a symmetric non-negative function that gives an affinity score for each pair of volumes.
One common way of constructing affinities is to take a distance metric $d : X \times X \rightarrow \mathbb{R}$ and apply a Gaussian kernel with a suitably chosen width $\sigma$ to form the \textit{affinity matrix} $W \in \mathbb{R}^{n \times n}$
\begin{align} \label{eq:gaussian_kernel}
    W_{ij} = w(\x_i, \x_j) = \exp \left( - d(x_i, x_j)^2/(2\sigma^2) \right) .
\end{align}
The \textit{degree matrix} $D \in \mathbb{R}^{n \times n}$ is defined to be the diagonal matrix that satisfies $D_{ii} = \sum_{j=1}^n W_{ij}$.
We use the \textit{Coifman-Lafon normalized graph Laplacian} \cite{CoifmanLafon2006}, which converges to the Laplace-Beltrami operator, regardless of the sampling density.
To compute this, one first performs a two-sided normalization of the affinity matrix,
\(
    \widetilde{W} = D^{-1} W D^{-1}
\)
and then computes the random-walk Laplacian,
\(
    \mathcal{L} = \widetilde{D}^{-1} \widetilde{W},
\)
where $\widetilde{D}$ is the degree matrix for $\widetilde{W}$.
The random-walk Laplacian is similar to a positive semi-definite symmetric matrix and hence its eigenvectors are real and its eigenvalues are non-negative.
The all-ones vector is an eigenvector of $\mathcal{L}$ with eigenvalue zero \cite{VonLuxburg2007}. 
Let $\phi_0, \phi_1, \ldots, \phi_{n-1} \in \mathbb{R}^n$ be eigenvectors of $\mathcal{L}$ with corresponding eigenvalues
$ 0 = \lambda_0 \le \lambda_1 \le \ldots \le \lambda_{n-1}$.
We think of the eigenvectors $\phi_{\ell}$ as real-valued functions on $X$, by identifying $\phi_{\ell}(\x_i) = (\phi_{\ell})_i$.
\noindent The \textit{$k$-dimensional diffusion map} $\Psi_t^{(k)}:X \to \mathbb{R}^k$ is defined by:
\begin{align*}
    \x_i \mapsto \big{(} \lambda_1^t \phi_{1}(\x_i), \ldots, \lambda_k^t \phi_{k}(\x_i) \big{)}.
\end{align*}
The mapping $\Psi_t^{(k)}$ gives a system of $k$ coordinates on $X$, which captures the intrinsic geometry of $\mathcal{M}$.
In our simulations, we used $t=0$, in which case diffusion maps coincide with \textit{Laplacian eigenmaps} \cite{BelkinNiyogi2003}.

The diffusion map depends on the choice of affinity.
The typical choice is a Gaussian kernel as defined in Eq.~\eqref{eq:gaussian_kernel} that is based on a Euclidean (or $\ell_2$) distance function,

$$d_{\ell_{2}}(\x_i, \x_j) = \Vert \x_i - \x_j \Vert_{2}.$$
We propose instead to base the Gaussian kernel of Eq.~\eqref{eq:gaussian_kernel} on the \textit{Earthmover's distance (EMD)}, also known as the \textit{Wasserstein metric} \cite{Villani2009}.
EMD is popular in various applications, e.g., image retrieval \cite{RubnerTomasiGuibas2000},
however, to the best of our knowledge, it has never been used to define affinities for manifold learning algorithms.
To define this distance, consider two 3-D density maps $\x_i, \x_j \in \mathbb{R}^{L^3}$ that are non-negative and normalized to unit mass.
These densities define probability measures on the set of voxels, $[L]^3$, where $[L] = \{1, \ldots, L\}$.
We set:
$$d_{\textup{EMD}}(\x_i, \x_j) \, = \, \min_{\pi \in \Pi(\x_i, \x_j)} \,\, \sum_{\u \in [L]^3} \sum_{\v \in [L]^3} \pi(\u, \v) \Vert \u - \v \Vert_{2}, $$
where $\Pi(\x_i, \x_j)$ is the set of joint probability measures on $[L]^3 \times [L]^3$ with marginals $\x_i$ and $\x_j$, respectively.

Mathematically, EMD amounts to a linear program in $\mathcal{O}(L^6)$ variables subject to $\mathcal{O}(L^3)$ constraints, i.e., a significant computation. 
However, in the wavelet domain \cite{Mallat1999}, EMD enjoys a
fast (weighted-$\ell_1$) wavelet approximation \cite{ShirdhonkarJacobs2008}, which we refer to as WEMD: %\cite{ShirdhonkarJacobs2008} (accurate up to known multiplicative constants):
\begin{equation}\label{eq:wavelets}
d_{\textup{WEMD}}(\x_i,\x_j) = \sum_{\lambda} 2^{-5s/2} \, \lvert \mathcal{W}\x_i(\lambda) - \mathcal{W}\x_j(\lambda) \rvert.
\end{equation}
Here, $\mathcal{W}\x$ denotes the \textit{3-D wavelet transform} of $\x$, and the index $\lambda$ contains the shifts $(m_1, m_2, m_3) \in \mathbb{Z}^3$ and the scale $s \in \mathbb{Z}_{\ge 0}$.  More explicitly, $\mathcal{W}$ decomposes $\x = \x[u_1,u_2,u_3]$ with respect to an orthonormal basis of functions, 
\begin{equation*}
    2^{3s/2} f(2^s u_1 - m_1) \, g(2^s u_2 - m_2) \, h(2^s u_3 - m_3), 
\end{equation*}
for varying $s \in \mathbb{Z}_{\ge 0}$, varying $(m_1, m_2, m_3) \in \mathbb{Z}^3$, and  $(f,g,h)$ ranging over $\{\psi, \omega\}^{3} \setminus \{(\omega, \omega, \omega)\}$ where
$\psi, \omega$ are certain  1-D functions called the \textit{mother and father wavelet} \cite{Mallat1999}.
Formula \eqref{eq:wavelets}  approximates  EMD in the sense that $d_{\textup{EMD}}$ and $d_{\textup{WEMD}}$ are strongly equivalent metrics, i.e., there exist constants $C \ge c > 0$ such that for all $\x, \y \in 
\mathbb{R}^{L^3}$, we have:
$$ c \cdot d_{\textup{WEMD}}(\x,\y) \, \le \, d_{\textup{EMD}}(\x,\y) \, \le \, C \cdot d_{\textup{WEMD}}(\x,\y).$$
Moreover, there are known bounds on the ratio $C / c$, depending on the type of wavelet used.
We have chosen the Coiflets~3 wavelet since it gives a small ratio \cite{ShirdhonkarJacobs2008}.
Wavelet transforms are computed in linear time, thus the same holds for the EMD approximation.
We 
 implemented the approximation \eqref{eq:wavelets}, using the PyWavelets package \cite{Lee2019}.
We computed the wavelet transform up to scale $s = 5$ for accurate truncation of Eq.
\eqref{eq:wavelets}.
This overparameterizes the volumes by a factor of $\approx 3$.

\section{Results}

\begin{figure}[t]
    \centering
    \includegraphics[width=0.99\linewidth]{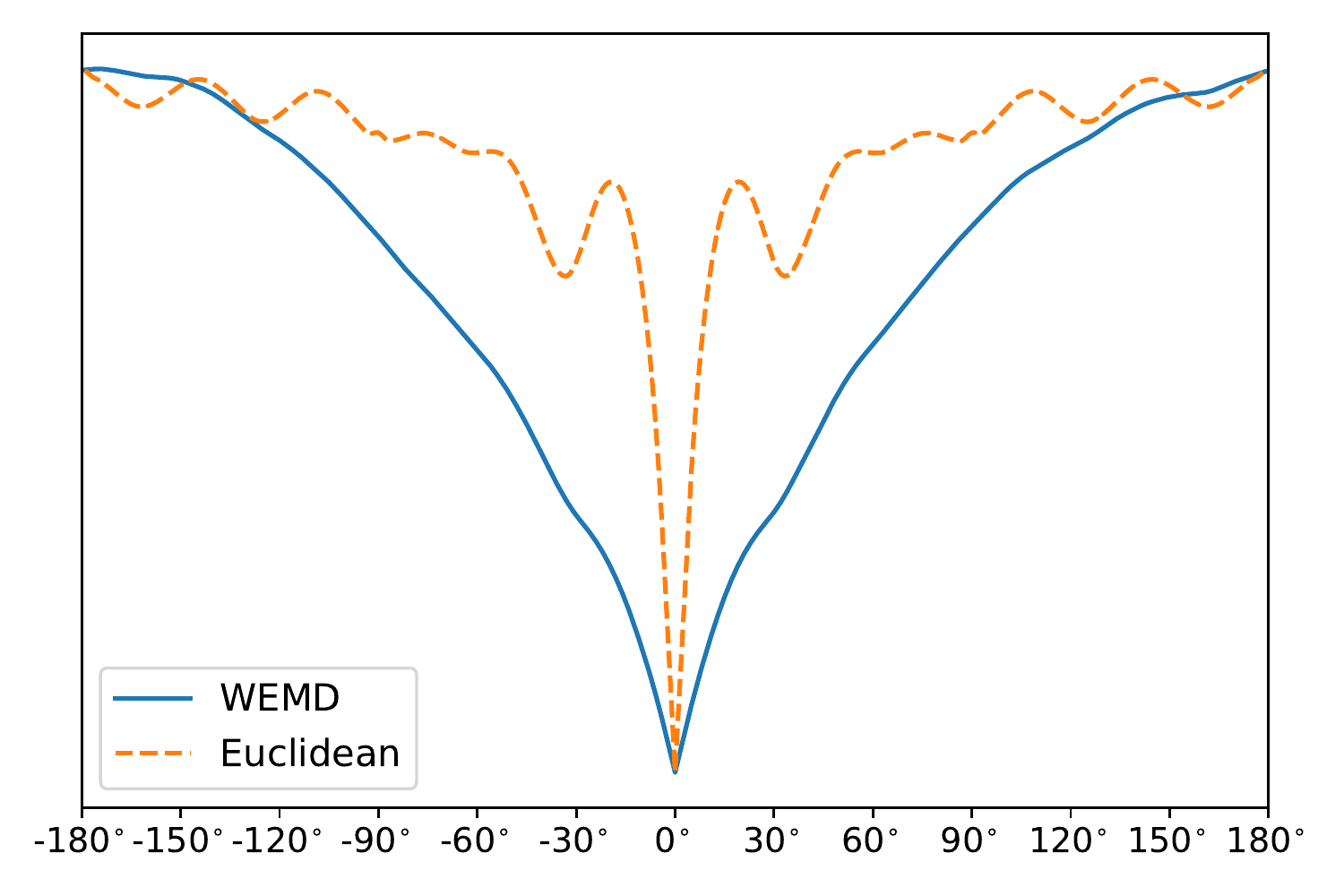}
    \caption{\textit{Euclidean distance vs. WEMD} as functions of the angle between two angles of the ATP synthase
rotor (see Fig.~\ref{fig:atp_synthase}). The $\ell_2$ distances are scaled to be comparable to the WEMD. The WEMD is monotone
in the magnitude of the angular difference for almost the entire range whereas the Euclidean distance exhibits this behavior only up to
about $\pm19\degree$.}
    \label{fig:compare_angle}
\end{figure}

\begin{figure}[t]
\begin{tabular}{llll}
    \toprule
    $n$ & Wavelet transform & WEMD distances & $\ell_2$ distances \\
    \midrule
    25 & 60 & 0.9 & 0.8\\
    50 & 121 & 3.4 & 1.2\\
    100 & 243 & 13 & 2.4\\
    200 & 481 & 50 & 4.8 \\
    400 & 965 & 221 & 9.4\\
    800 & 1932 & 844 & 20.6\\
    \bottomrule
\end{tabular}
\caption{\textit{Running times [sec]} for computing the fast wavelet transform, all pairwise wavelet Earthmover approximations and all pairwise $\ell_2$ distances.}
\label{fig:runtime}
\end{figure} 

\begin{figure*}[t!]
    \centering
    
    \begin{tabular}{ccccccc}
        \begin{tabular}{c} Noiseless\\Euclidean\\\\\\\\\\ \end{tabular}  & \includegraphics[height=0.80in]{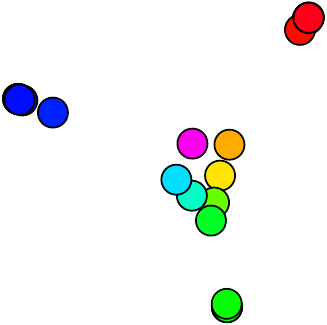} & \includegraphics[height=0.80in]{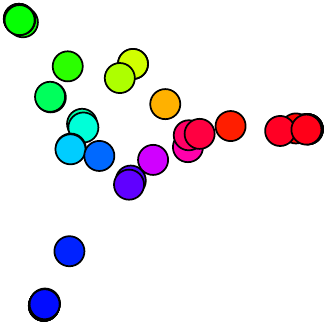} & \includegraphics[height=0.80in]{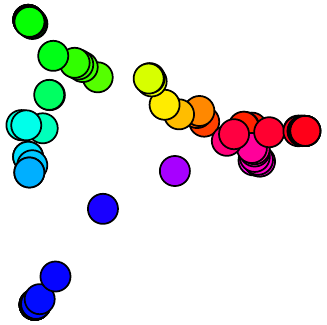} & \includegraphics[height=0.80in]{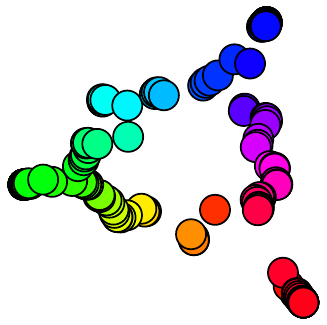} & \includegraphics[height=0.80in]{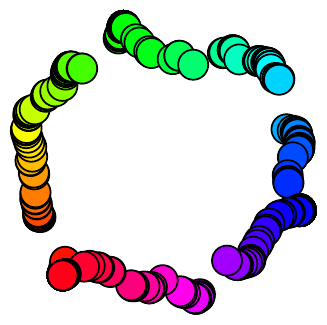} & \includegraphics[height=0.80in]{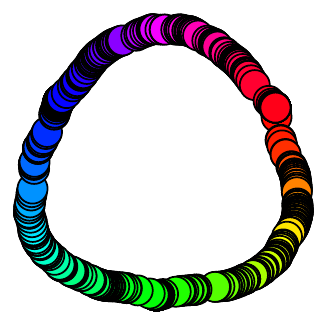} \\
        \\[-1.2cm]

        \begin{tabular}{c}Noiseless\\WEMD\\\\\\\\\\\end{tabular} & \includegraphics[height=0.80in]{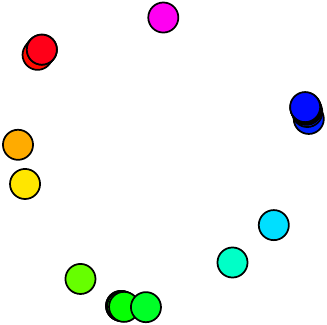} & \includegraphics[height=0.80in]{Images/emd_embedding_CL_25_noiseless_2019} & \includegraphics[height=0.80in]{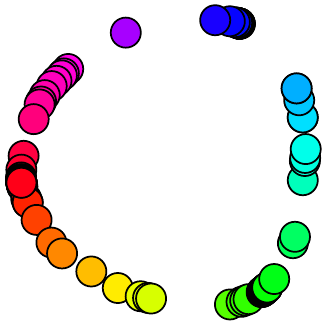} & \includegraphics[height=0.80in]{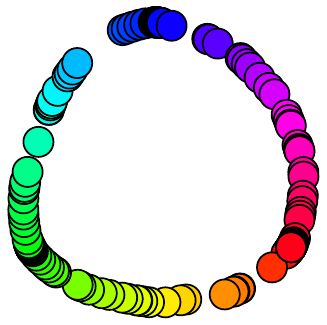} & \includegraphics[height=0.80in]{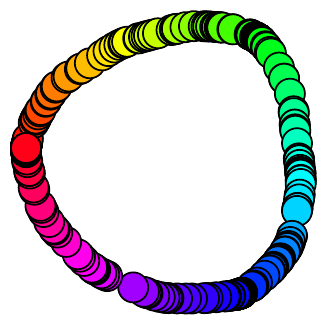} & \includegraphics[height=0.80in]{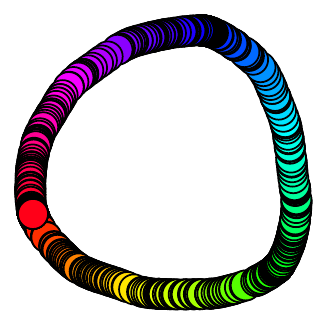}  \\
        \\[-1.2cm]

        \midrule
        \\[-0.1cm]
        \begin{tabular}{c}Noisy\\Euclidean\\\\\\\\\\ \end{tabular} & \includegraphics[height=0.80in]{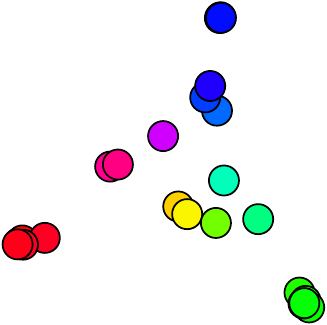} & \includegraphics[height=0.80in]{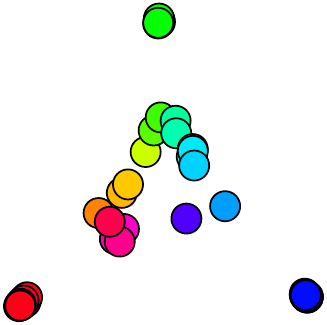} & \includegraphics[height=0.80in]{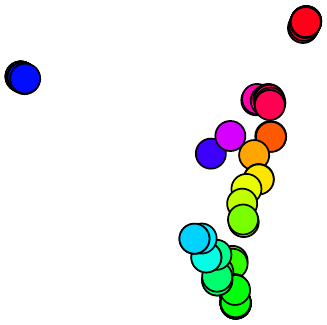} & \includegraphics[height=0.80in]{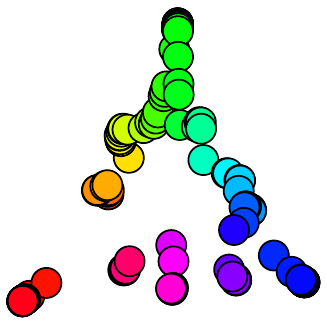} & \includegraphics[height=0.80in]{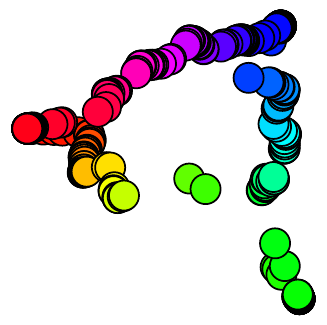} & \includegraphics[height=0.80in]{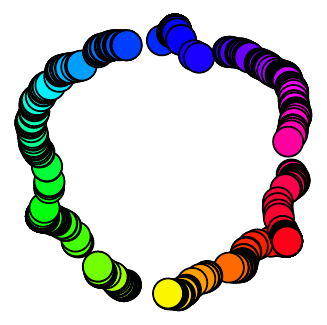}\\

        \\[-1.2cm]

        \begin{tabular}{c}Noisy\\WEMD\\\\\\\\\\ \end{tabular} & \includegraphics[height=0.80in]{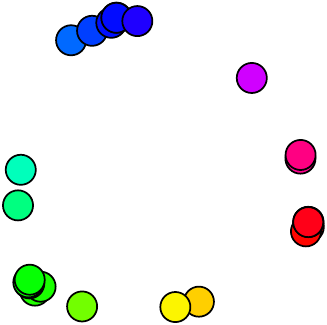} & \includegraphics[height=0.80in]{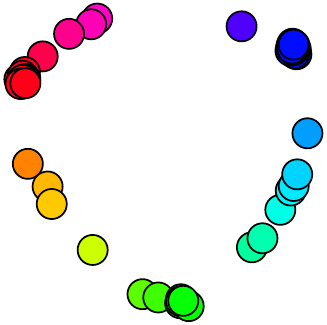} & \includegraphics[height=0.80in]{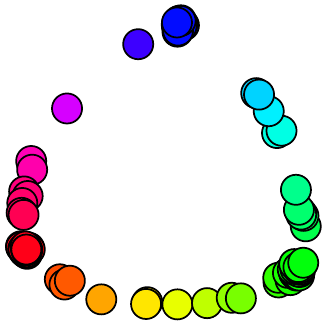} & \includegraphics[height=0.80in]{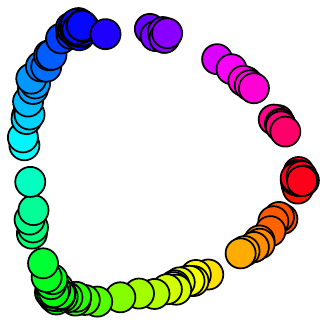} & \includegraphics[height=0.80in]{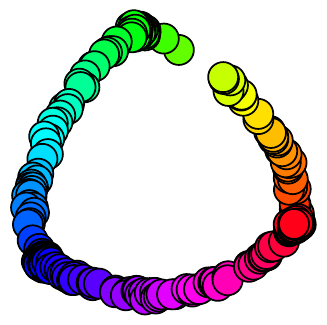} & \includegraphics[height=0.80in]{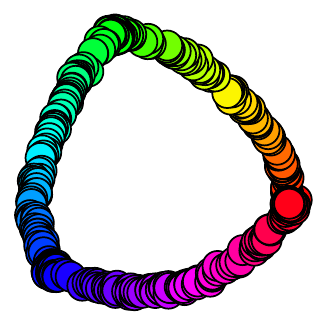}\\
        \\[-1.2cm]
        $n$ & 25 & 50 & 100 & 200 & 400 & 800
    \end{tabular}

    \caption{\textit{Main results.} Euclidean vs. EMD-based diffusion mappings on the clean and noisy ATP synthase datasets for sample sizes $n=25,50,100,200,400,800$.
    The Euclidean diffusion maps need more than $400$ samples to capture the intrinsic geometry whereas WEMD manages to do so with merely $n=25$ samples. The colors encode the (ground truth) angle.}
    \label{fig:embeddings}
\end{figure*}

To test our methods, we generated two synthetic datasets of 3\nobreakdash-D density maps that are
simplified models of the conformation space of ATP synthase \cite{YoshidaMuneyukiHisabori2001}.
This enzyme is a molecular stepper motor with a central asymmetric axle that rotates in steps of 120$\degree$ relative to the F$_1$ subunit, with short transient motion in-between the three dominant conformations.
Here, the intrinsic geometry is a circle, with a sampling density concentrated around three equispaced angles. 
We simulated this motion by generating 3-D density maps in which the F$_1$ subunit is held in place while the F$_0$ and axle subunits are rotated together by a random angle.
The angles were drawn i.i.d. according to the following mixture model:
\begin{align*}
    \frac{2}{5} U[0,360] + \frac{1}{5} \mathcal{N}(0,1) + \frac{1}{5} \mathcal{N}(120,1) + \frac{1}{5} \mathcal{N}(240,1),
\end{align*}
where $U$ and $\mathcal{N}$ denote uniform and Gaussian distributions, respectively.
To form our datasets, we downloaded entry 1QO1 \cite{Stock1999} from the Protein Data Bank \cite{Roseetal.2017}, produced 3\nobreakdash-D density maps at a 6\AA \ resolution with array dimensions $47 \times 47 \times 107$ using the \texttt{molmap} command in UCSF Chimera \cite{Chimera2004}, and then took random rotations of the F$_0$ and axle subunits.
From this, we generated a clean dataset and a noisy dataset.
For the latter, i.i.d. Gaussian noise was added with mean zero and standard deviation equal to one-tenth of the maximum voxel value.

We first tested the plausibility of our proposal by comparing the EMD approximation to the Euclidean distance for a range of angular differences using the noiseless dataset (Fig.~\ref{fig:compare_angle}). 
We then performed 2-dimensional diffusion maps for various sample sizes, using both the Euclidean distance and the wavelet-based approximation to the EMD, as described in the previous section.
The resulting embeddings are shown in Fig. \ref{fig:embeddings}.
The value of the width parameter $\sigma$ in the Gaussian kernel \eqref{eq:gaussian_kernel} was hand-picked to yield the best results.
We note that for the Euclidean diffusion maps, careful tuning of $\sigma$ was required.
However, this was not necessary for the EMD approximation, where a wide range of $\sigma$ values gave excellent results. Running times (on an Intel Core i7) for the computation of EMD and Euclidean-based diffusion maps are listed in Fig.~\ref{fig:runtime}. 

\section{Conclusion}

In this paper, we proposed to use Earthmover-based affinities in the diffusion maps framework to analyze molecular conformation spaces.  We showed that this results in a marked decrease in the number of samples needed to capture the intrinsic conformation space of ATP synthase.
The method is computationally tractable, thanks to a fast wavelet approximation, and robust to noise. Our results show promise, particularly for the analysis of cryo-EM datasets with continuous heterogeneity.
More broadly, EMD-based manifold learning could be applied to analyze the variability of other collections of 3-D shapes \cite{Ovsjanikov2011}, 2-D images \cite{RubnerTomasiGuibas2000}, videos and other signals, e.g., to better model animal motion \cite{Hu2009}.
Our work also raises several interesting theoretical questions: in which cases can one prove that EMD-based manifold learning has a lower sample complexity than manifold learning based on the Euclidean distance?  
More ambitiously, are there reasonable generative models for variability where EMD is the optimal distance metric?

\section{Reproducibility}

Code for reproducing the results in this paper is available at
\noindent \href{http://github.com/nathanzelesko/earthmover}{http://github.com/nathanzelesko/earthmover}

\section{Acknowledgments}
The authors thank Ariel Goldstein, William Leeb, Nicholas Marshall and Stefan Steinerberger for interesting discussions.
This work was supported in parts by AFOSR FA9550-17-1-0291, ARO W911NF-17-1-0512, the Simons Collaboration in Algorithms and Geometry, the Simons Investigator Award, the Moore Foundation Data-Driven Discovery Investigator Award and NSF BIGDATA Award IIS-1837992.

\newpage

\bibliographystyle{IEEEbibWithDOI}
\bibliography{earthmover}

\end{document}